\documentclass{optica-article}

\journal{opticajournal} % for journals or Optica Open

\articletype{Research Article}

\usepackage[capitalise]{cleveref}
\usepackage{tabularx,booktabs}

\usepackage{physics}
\usepackage{dsfont}

\usepackage{lineno}
\usepackage[utf8]{inputenc}
\newcommand{\figpanel}[2]{Fig.~\hyperref[#1]{\ref*{#1}(#2)}}

\newcommand{\inlinemel}[3]{\langle#1\vert#2\vert#3\rangle}

\begin{document} 
\title{Nonperturbative cavity quantum electrodynamics: is the Jaynes-Cummings model still relevant?}

\author{Daniele De Bernardis,\authormark{1, *} Alberto Mercurio,\authormark{2,3} and Simone De Liberato\authormark{4,5}}

\address{\authormark{1}National Institute of Optics [(Consiglio Nazionale delle Ricerche CNR)–INO)], care of European Laboratory for
Non-Linear Spectroscopy (LENS), Via Nello Carrara 1, Sesto Fiorentino, 50019, Italy\\
\authormark{2}Laboratory of Theoretical Physics of Nanosystems (LTPN), Institute of Physics, Ecole Polytechnique Fédérale de Lausanne (EPFL), CH-1015 Lausanne, Switzerland\\
\authormark{3}Center for Quantum Science and Engineering, EPFL, CH-1015 Lausanne, Switzerland\\
\authormark{4}School of Physics and Astronomy, University of Southampton, Southampton SO17 1BJ, UK\\
\authormark{5} Istituto di Fotonica e Nanotecnologie – Consiglio Nazionale delle Ricerche (CNR), Piazza Leonardo da Vinci 32, Milano, Italy
}

\email{\authormark{*}daniele.debernardis.89@gmail.com} %% email address is required; see note below about the corresponding author designation

% use {asbstract*} to suppress the copyright line. Copyright information will be added in production

\begin{abstract*}
In this tutorial review, we briefly discuss the role that the Jaynes-Cummings model occupies in present-day research in cavity quantum electrodynamics with a particular focus on the so-called ultrastrong coupling regime.
We start by critically analyzing the various approximations required to distill such a simple model from standard quantum electrodynamics. We then discuss how many of those approximations can, and often have been broken in recent experiments. The consequence of these failures has been the need to abandon the Jaynes-Cummings model for more complex models. In this, the quantum Rabi model has the most prominent role and we will rapidly survey its rich and peculiar phenomenology.
We conclude the paper by showing how the Jaynes-Cummings model still plays a crucial role even in non-perturbative light-matter coupling regimes.
\end{abstract*}

%%%%%%%%%%%%%%%%%%%%%%%%%%  body  %%%%%%%%%%%%%%%%%%%%%%%%%%
\section{Introduction}

The Jaynes-Cummings model (JCM) is a pivotal theoretical object in quantum optics, describing the quintessential interaction between light and matter at the quantum level. It models the simplest quantum emitter, a two-level system (TLS), interacting with a single electromagnetic degree of freedom, a discrete mode of a photonic cavity. Their interaction is described by a Hamiltonian composed of two terms with transparent physical interpretation: the first describes the emitter transitioning between the ground and the excited state by the absorption of a photon, and the second, its Hermitic conjugate, describes its de-excitation caused by photon emission.
This simplicity has made the JCM an outstanding pedagogical tool in quantum optics, and a fundamental framework for understanding and analyzing light-matter interactions in cavity quantum electrodynamics (CQED) systems. Since the Nobel-worth experiments of Haroche and Wineland \cite{haroche_nobel_2013, wineland_nobel_2013}, which for the first time allowed to experimentally measure some of the most peculiar predictions of the JCM, the study of CQED has grown into one of the most active in physics, with impact in fields as different as quantum information \cite{zasedatelev_room-temperature_2019,berloff_realizing_2017, ballarini_polaritonic_2020}, chemistry \cite{schwartz_reversible_2011,cwik_excitonic_2016,nagarajan_chemistry_2021}, photonics \cite{nomura_laser_2010,delteil_towards_2019}, material engineering \cite{brodbeck_experimental_2017,cortese_excitons_2021,appugliese_breakdown_2022,garcia-vidal_manipulating_2021, de_bernardis_magnetic-field-induced_2022}, and many others. As the boundary of knowledge was pushed forward, some of the underlying hypotheses that allowed to distilling of the Platonic simplicity of the JCM out of the complexity of an interacting light-matter system have been stretched or altogether broken. This led in turn to a vast theoretical effort to extend the JCM to understand and solve these shortcomings.

This tutorial review aims to give an overview of the physics of and beyond the JCM, exploring under which conditions different approximations break down, how the theory can be modified to accommodate those situations, and which novel phenomenology becomes observable. 

\section{The minimal description of a cavity QED system}
\label{sec: Derivation of the JCM}

\subsection{Introducing the Jaynes-Cummings model}
\begin{figure}[t]
    \centering
    \includegraphics{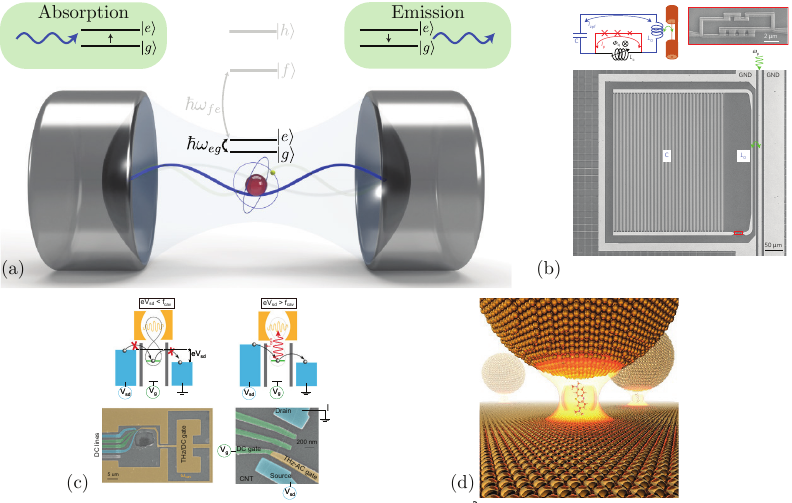}
    \caption{(a) Schematic illustration of the quantum light-matter interaction in a Fabry-P\'erot cavity. A single atom is coupled to a single mode of the cavity field (blue wave). The figure shows the four assumptions that lead to the Quantum Rabi model. (i) A single atom is present in the cavity. (ii) Single-mode approximation, which assumes that only one cavity mode is relevant for the atom-field coupling. The other cavity modes are shaded. (iii) Dipolar approximation, with a constant electromagnetic field across the atom region. (iv) The two-level approximation, 
    neglecting the higher-lying states shaded in the figure. 
    The rotating wave approximation then leads to the Jaynes-Cumming model, whose two interaction terms are depicted on the top of the mirrors. These processes result in the coherent exchange of energy between the atom and the field. (b) A superconducting circuit that reaches the USC regime, taken from Ref.~\cite{yoshihara_superconducting_2016}. (c) Quantum dot coupled to a THz cavity in the USC regime, taken from Ref.~\cite{valmorra_vacuum-field-induced_2021}. (d) Plasmonic nanocavity device, taken from Ref.~\cite{chikkaraddy_single-molecule_2016}.
    }
    \label{fig:3d-figure}
\end{figure}

The JCM model describes the idealized system represented in Fig. \ref{fig:3d-figure}: a TLS coupled to a single electromagnetic mode of a cavity.
Its Hamiltonian reads
\begin{equation}\label{eq:JC_Int_ham}
    \hat{H}_{\rm JC} = \hbar \omega_c \hat{a}^{\dag} \hat{a} + \frac{\hbar\omega_{eg}}{2} \hat{\sigma}_z + \hbar \Omega_{\rm R} \left( \hat{a} \hat{\sigma}_+ + \hat{a}^{\dag}\hat{\sigma}_- \right).
\end{equation}
Here $\hat{\sigma}_z$ and $\hat{\sigma}_-=\hat{\sigma}_+^{\dag}=\hat{\sigma}_x - i \hat{\sigma}_y$ are Pauli matrices, $\hat{a}$ is the annihilation operator of a cavity photon with frequency $\omega_c$, satisfying $[\hat{a}, \hat{a}^{\dag}]=1$, $\omega_{eg}$ is the frequency of an optically active transition, and $\Omega_{\rm R}$ is the so-called vacuum Rabi frequency, which quantifies the light-matter interaction strength. 

The intuitive understanding of light-matter interactions provided by the JCM is based on the concepts of absorption and emission, spontaneous and stimulated, first introduced by Einstein through his \emph{A, B coefficients} \cite{einstein_volume_nodate}.
The interaction part of the JCM Hamiltonian in Eq. \eqref{eq:JC_Int_ham}, proportional to the vacuum Rabi frequency $\Omega_{\text{R}}$, is of the flip-flop type and whenever a photon is destroyed the TLS is excited, and vice versa.
As a consequence, the sum of photon number plus TLS excitation is a conserved quantity, which can be formalized by introducing the excitation number operator
\begin{equation}
  \hat N = \hat{a}^{\dag}\hat{a} + \frac{1}{2}\left[\hat{\sigma}_z + 1\right],
\end{equation}
that commutes with the JCM Hamiltonian $[\hat N, \hat{H}_{\rm JC}]=0$.
In the spirit of the Noether theorem, conservation laws are linked to symmetries. 
So the conservation of the excitation number is reflected in the continuous $U(1)$ symmetry of the JCM where $\hat N$ is the generator of the phase shift transformation
\begin{equation}
    \begin{split}
        e^{i\theta \hat N}\hat{a}e^{-i\theta \hat N} = \hat{a} e^{i\theta},
        \quad 
        e^{i\theta \hat N}\hat{\sigma}_-e^{-i\theta \hat N} = \hat{\sigma}_- e^{i\theta},
    \end{split}
\end{equation}
that leaves the Hamiltonian in Eq. \eqref{eq:JC_Int_ham} invariant.
The symmetry and the commutativity of $\hat N$ with $\hat H$ allow us to express the JCM Hamiltonian in Eq. \eqref{eq:JC_Int_ham} in block-diagonal form, where each block is characterized by the quantum number $n\in\mathbb{N}_+$.
Since each $n\neq0$ eigenvalue of $\hat N$ is doubly-degenerate, with eigenstates $|n, g\rangle \, , |n-1, e\rangle $, the JCM can be put in two-by-two block-diagonal form (one-by-one for the $n=0$ non-degenerate ground state $|0,g\rangle$)
\begin{align}\label{eq:two_by_two_JCM_matrix}
\hat{H}^{0}_{JC}&=-\frac{\hbar\omega_{eg}}{2},\\
\hat{H}^{n}_{JC}&=\left[\begin{array}{cc}
         \hbar \omega_c n - \frac{\hbar \omega_{eg}}{2}  & \hbar\Omega_{\text{R}}\sqrt{n} \\
        \hbar \Omega_{\text{R}}\sqrt{n} & \hbar\omega_c (n-1) +\frac{\hbar\omega_{eg}}{2}
    \end{array}
    \right],
\end{align}

where for each $n>0$ the eigenvalues of the n$^{\text{th}}$ block, $\omega_{n,\pm}$, describe dressed states with $n$ excitations.
{\color{black}The JCM spectrum is then immediately given by $\omega_{n,\pm} = -\omega_{eg}/2 + \omega_c n - \Delta/2 \pm \sqrt{\Delta^2/4 + \Omega_{\text{R}}^2 n}$, where we have introduced the cavity-TLS detuning $\Delta = \omega_c - \omega_{eg}$.}
The saturation of the TLS makes the system nonlinear, leading to a $n$-dependent intra-doublet splitting which at resonance{\color{black}, $\Delta=0$,} reads
\cite{larson_jaynescummings_2021, haroche_exploring_2013}
\begin{equation}
\label{splitting}
    \omega_{n, +} - \omega_{n, -} = 2\Omega_{\rm R}\sqrt{n}.
\end{equation}
{\color{black}The JCM nonlinearity is at the base of many important effects in quantum optics, among all it is worth mentioning the photon blockade \cite{tian_quantum_1992, rebic_large_1999, brecha_n_1999, birnbaum_photon_2005}, which is a central object in the developments of quantum technology \cite{kimble_quantum_2008,kurizki_quantum_2015}.

}

Notice that when the vacuum Rabi frequency is larger than the bare excitation frequencies $\Omega_{\rm R} > \omega_c,\omega_{eg}$ the lowest eigenenergy of the JCM becomes negative, replacing the ground state of the system and thus changing its equilibrium properties.
This prediction gives us a hint regarding the potential interest in studying the regime in which the vacuum Rabi frequency is comparable to or larger than the bare cavity and TLS frequencies. 

\subsection{Why we need to go beyond the JCM: three experimental examples}

The JCM is usually employed to describe systems in the \emph{strong coupling} regime of CQED \cite{kimble_cavity_2003}, in which the vacuum Rabi frequency becomes larger than the loss rates for the light ($\gamma$ ) and matter ($\kappa$) degrees of freedom, and we can resolve the resonant $n=1$ splitting $\omega_{1,+}-\omega_{1,-}=2\Omega_{\mathrm{R}}$ as from Eq.\ref{splitting}.
It is thus useful to introduce the cooperativity parameter $C=4\Omega_{\rm R}^2/(\gamma \kappa)$\cite{haroche_exploring_2013}, with $C>1$ marking the onset of the strong coupling regime.

When the coupling becomes instead comparable to the frequencies of the bare excitations, the system enters a different regime characterized by a novel non-perturbative phenomenology. Such a regime has been named ultrastrong coupling (USC) regime \cite{forn-diaz_ultrastrong_2019, frisk_kockum_ultrastrong_2019, le_boite_theoretical_2020}.
By analogy with the cooperativity, it is useful to introduce the normalized coupling parameter $\zeta = 4\Omega_{\rm R}^2/(\omega_c \omega_{eg})$, with $\zeta \sim O(1)$ identifying the USC regime \cite{rossatto_spectral_2017} (the value $\zeta=0.04$, corresponding in the resonant case to $\Omega_{\rm R}=0.1\omega_c$, is usually used as threshold but this is only a historical accident \cite{anappara_signatures_2009}).

In the state-of-the-art experiments with Rydberg atoms in high-finesse optical cavities $C \sim 10-100 \gg 1$ , but at the same time $\zeta \sim 10^{-5} \ll 1$~\cite{haroche_exploring_2013}.
%The main difficulty lies in the very principles of QED, which is by definition a perturbative theory due to the small value of its fundamental coupling strength: the fine structure constant $\alpha_{\rm fs} \approx {1} / {137}$.
% \begin{equation}
%     \alpha \approx \frac{1}{137}.
% \end{equation}
It was indeed shown in \cite{devoret_circuit-QED_2007} that for a single hydrogenoid atom coupled to a resonant electric field, the normalized coupling can be written as $\Omega_{\text{R}} / \omega_c \approx \alpha_{\rm fs}^{3/2} / (\ell \pi\sqrt{V})$, 
% \begin{align}
% \frac{\Omega_{\text{R}}}{\omega_c}\approx\frac{\alpha^{3/2}}{n \pi\sqrt{V}}, 
% \end{align}
where $\alpha_{\rm fs} \approx 1/137$ is the fine structure constant, $\ell$ is the principal quantum number, and $V$ the cavity volume expressed in units of a cube with half-wavelength sides. 
Without extreme subwavelength confinement, which is often accompanied by large losses \cite{khurgin_how_2015}, it is thus impossible to achieve non-perturbative light-matter USC on a single atom due to the fundamental hard bound imposed by the fine structure constant.
From this analysis, it seems that the single particle USC regime is forbidden by the very basic principles of QED.

To circumvent this bound one can only rely on artificial, highly engineered systems where the dynamics of the electromagnetic field is mediated by a material component \cite{devoret_circuit-QED_2007}.
In particular, considering superconducting circuits it was found that the fine structure constant is rescaled by the circuit impedance $\alpha_{\rm fs} \longmapsto \left(Z / Z_0\right)^{\ell}\alpha_{\rm fs}$,
% \begin{equation}
%     \alpha \longmapsto \left(\frac{Z}{Z_0}\right)^{\ell}\alpha, 
% \end{equation}
where $Z_0 = 1/(\epsilon_0c)$ is the vacuum impedance, while $\ell = \pm 1$ dependently from the origin of the coupling (capacitive, inductive,etc...)\cite{niemczyk_circuit_2010, jaako_ultrastrong-coupling_2016}.
Beyond superconducting circuits only, this concept has a more broad application and similar scaling can be found also for plasmonic cavities, metamaterials and in general for any sub-wavelength resonant structure \cite{de_bernardis_cavity_2018, saez-blazquez_can_2023}.
In such setups the impedance has no bound in principle (if not merely technological), and a fully non-perturbative USC is possible, with values of $\zeta\sim 0.01 - 100$.

In Table \ref{tab:exp_params_single_particle} we report three examples of CQED setups working in various frequency regimes where a TLS coupled to a photonic resonator reaches the USC regime.
\begin{table}[]
    \begin{tabularx}{\textwidth}{X|X|X|X|X}
         {\rm \bf System} & {\rm \bf Cavity $\omega_c$} & {\rm \bf Atom  $\omega_{eg}$} & {\rm \bf Coupling $\Omega_{\rm R}$} & $\zeta$  \\
         {\rm Superconducting circuits} \cite{yoshihara_superconducting_2016} & 35.2~{\rm GHz} & 23.9~{\rm GHz} & 35.2~{\rm GHz} & 6  \\
         {\rm Molecular plasmonic cavities} \cite{chikkaraddy_single-molecule_2016} & 452~{\rm THz} & 452~{\rm THz} & 73~{\rm THz} & 0.03  \\
         {\rm Graphene quantum dots} \cite{valmorra_vacuum-field-induced_2021} & 25~{\rm THz} & 3.8~{\rm THz} & 49~{\rm THz} & 101  \\
    \end{tabularx}
    \caption{Parameters of CQED setups in which a TLS coupled to a photonic resonator reaches the USC regime.}
    \label{tab:exp_params_single_particle}
\end{table}
Note that the performance of the molecular plasmonic setup \cite{chikkaraddy_single-molecule_2016} was just close to  USC physics. It was nevertheless shown how small modifications \cite{kuisma_ultrastrong_2022} could increase the coupling even further. 
In Fig. \ref{fig:3d-figure} we shows these three setups:
superconducting circuits (b) \cite{yoshihara_superconducting_2016} (also in their multimode or multi-qubit USC versions \cite{forn-diaz_ultrastrong_2017,wang_strong_2024,mehta_down-conversion_2023,tomonaga_one_2024, vrajitoarea_ultrastrong_2024}), carbon nanotube quantum dots in THz cavities (c) \cite{valmorra_vacuum-field-induced_2021}, and molecules in plasmonic resonators (d) \cite{chikkaraddy_single-molecule_2016,benz_single-molecule_2016,kuisma_ultrastrong_2022,jakob_giant_2023}.
Notwithstanding their differences, these USC systems share the same underlying CQED structure: a discrete electronic transition interacts coherently with a confined electromagnetic field. Still, the JCM does not correctly reproduce their features.

\section{Understanding the Jaynes-Cummings model from QED}
\label{sec:understanding_JCM}
In order to understand this failure, and obtain a new model applicable in the USC regime and which recovers the JCM for weaker coupling strengths, we need first to understand how the JCM itself is obtained.
Its clear depiction of light-matter interactions can be rigorously derived via a series of approximations from the non-relativistic quantum electrodynamics (QED) Hamiltonian. 
We list here the main steps to distill the JCM from the underlying QED theory, with the main approximations schematically represented in \cref{fig:3d-figure}.

%\begin{enumerate}

    %\item \emph{Dipolar approximation}
\subsection{Dipolar approximation}
        To start, the JCM considers the light-matter interaction in the dipolar approximation. Formally this is obtained by expressing the full non-relativistic QED Hamiltonian in the so-called Poincar\'e gauge \cite{cohen-tannoudji_lagrangian_1997}: truncating its multipolar expansion to the lowest order, one obtains the dipole gauge Hamiltonian \cite{stokes_identification_2021}. In simpler terms, when the electromagnetic field does not vary too much on the length scales of the TLS spatial extension, it can be considered spatially uniform. 
        As a consequence, the light-matter interaction Hamiltonian can be derived by considering the energy of {\color{black}an electric dipole $\hat{\mathbf{d}}$ in a uniform electric field $\hat{\mathbf{E}}$ or a magnetic dipole $\hat{\mathbf{m}}$ in a uniform magnetic field $\hat{\mathbf{B}}$ 
        \begin{equation}
            \label{eq: general form of d * E interaction}
            \hat{H}_I \propto \hat{\mathbf{d}}\cdot \hat{\mathbf{E}} \quad {\rm or} \quad \hat{\mathbf{m}} \cdot \hat{\mathbf{B}}.
        \end{equation}}
        {\color{black} While standard cavity QED discussed in textbook is typically due to the electric dipole coupling \cite{cohentannoudji_atomphoton_1998,milonni_quantum_1994,haroche_exploring_2013}, it is worth stressing that a magnetic coupling leads to the very same phenomenology either through the Zeeman term \cite{imamoglu_cavity_2009, jenkins_coupling_2013, jenkins_scalable_2016, martinez-perez_quantum_2019, roman-roche_photon_2021, rollano_high_2022} or through orbital magnetism \cite{andolina_theory_2020, bacciconi_first-order_2023, mercurio_photon_2024}. }
        
        This is a safe approximation for atoms in microwave cavities \cite{haroche_nobel_2013}, but much less so for
        extended objects like molecules or quantum dots \cite{craig_molecular_1998, andersen_strongly_2011, tighineanu_accessing_2014}, or in nanophotonic resonators in which higher-order modes can be excited \cite{raza_multipole_2015} and can lead to the phenomenon of fluorescence quenching \cite{anger_enhancement_2006}.
      In general, the presence of selection rules can suppress the dipole coupling in favor of other type of multipolar interactions due to symmetry \cite{craig_molecular_1998, felicetti_ultrastrong-coupling_2018, koski_strong_2020}. 

\subsection{Modelling the emitter as a TLS}
        While transitions in spin doublets can be exactly modeled as TLS, most quantum emitters are implemented with electronic transitions. The number of trapped electronic states can be substantially larger than two, and the possibility of focusing only on a single, discrete transition, between the ground state ($|g\rangle$) and a single excited state ($|e\rangle$) rests on the assumption that the coupling with all the other parts of the spectrum can be neglected. The evolution of the system can thus be considered to span only the two-dimensional Hilbert space $\left[\ket{g},\ket{e}\right]$. 
        Using the Pauli matrices we can then describe any operator in the two-level subspace, for instance, the atomic {\color{black}electric} dipole operator becomes
        \begin{equation}
        \hat{\mathbf{d}}\propto |e\rangle \langle g| + |g\rangle \langle e| = \hat{\sigma}_- + \hat{\sigma}_+.
        \end{equation}
        The reason to ignore the other energy levels is that they are {\it out-of-resonance}, a justification so pervasive in physics that its assumptions are sometimes overlooked and can become surprisingly fragile in state-of-the-art CQED setups \cite{de_bernardis_breakdown_2018}. When this approximation breaks down modifications of the electronic wavefunction can be obtained. The coupled wavefunctions are interference patterns of the bare ones, and thus very sensitive and tunable \cite{khurgin_excitonic_2001,askenazi_ultra-strong_2014,cortese_strong_2019,levinsen_microscopic_2019}, a phenomenology referred to as very strong coupling.

    %\item \emph{Considering a single cavity mode}
\subsection{Considering a single cavity mode}
        Analogously to the matter degrees of freedom, the photonic resonator also hosts a complete set of modes \cite{gubbin_real-space_2016}, both discrete and belonging to a continuum. 
       If the field is strongly confined in a cavity or in any artificial structure (e.g. subwavelength resonators), the energy spacing between the different electromagnetic modes is such that we can discard all of them except the one that is most resonant with the optically active transition of interest.
        In such a way the cavity dynamics is completely described as a harmonic oscillator, with annihilation operator $\hat{a}$, and the electric {\color{black} and magnetic} field operators are given by{\color{black}
        \begin{equation}
            \hat{\mathbf{E}} \propto \hat{a} + \hat{a}^{\dag}, \quad \hat{\mathbf{B}} \propto -i\left( \hat{a} - \hat{a}^{\dag}\right) .
        \end{equation}}
        When this approximation is violated and the light-matter coupling becomes larger than the free spectral range we reach a different regime which has been called superstrong coupling \cite{kuzmin_superstrong_2019}. In such a regime considering a single photonic mode allows faster-than-light signalling \cite{sanchez_munoz_resolution_2018}.
        
        Analogously to the electronic case in the very strong coupling regime, the electromagnetic field of the coupled modes in the superstrong coupling regime are linear superpositions of multiple uncoupled electromagnetic modes, and dynamical modifications of subwavelength mode profile can be achieved \cite{cortese_real-space_2023,mornhinweg_sculpting_2024}. 
        The coupling of different electromagnetic modes also provides extra degrees of freedom to the wavefunctions of the coupled light-matter eigenmodes, which at larger values of the normalized coupling can bend to avoid the dipoles. This also realizes one among the various mechanisms that leads to the phenomenon of light-matter decoupling \cite{de_liberato_light-matter_2014} described in Sec. \ref{SecDecoupling}.
        
    %\item \emph{Applying the rotating wave approximation}
\subsection{Applying the rotating wave approximation}
        The rotating wave approximation (RWA) is ubiquitous in physics even if known with different names, another one being secular approximation from its use in celestial mechanics, where the errors introduced would have become observable only over centuries. 
        %In light-matter interaction, it can be derived from the optical Bloch equation for a driven TLS \cite{cohentannoudji_atomphoton_1998}, and 
        In the context of CQED is implemented by the following reduction {\color{black}(the same holds for the magnetic coupling)}
        \begin{equation}
       \hat{\mathbf{d}}\cdot \hat{\mathbf{E}} \propto \left(\hat{\sigma}_- + \hat{\sigma}_+\right)\left( \hat{a} + \hat{a}^{\dag} \right)\approx \,\hat{\sigma}_-\hat{a}^{\dag} + \hat{\sigma}_+\hat{a}.
        \end{equation}
        This approximation    
        is the crucial one in defining the light-matter interactions in terms of absorption/emission processes, as it consists in neglecting terms which lack an intuitive physical understanding and whose impact becomes
        non-negligible only in the USC regime.
        
        Given the importance of this approximation for the definition of the ultrastrong coupling regime a more in-depth discussion of the consequences of going beyond the rotating-wave-approximation will be given in the next section (Sec. \ref{SecRWA}).

%\end{enumerate}

\section{Unwind the rotating-wave: the Rabi model}
\label{SecRWA}

For large enough values of the light-matter coupling strength all these approximation but the dipole one eventually break down \cite{de_bernardis_cavity_2018, le_boite_theoretical_2020}.
In the region of interest for current experiments, the RWA is the first one to be broken and hence the first we discuss here.

\subsection{Re-introducing the counter-rotating terms}

The JCM without the RWA it the so-called \emph{quantum Rabi model} (QRM), described by the Hamiltonian
\begin{equation}\label{eq:ham_Rabi}
    \hat{H}_{\rm R} = \hbar\omega_c \hat{a}^{\dag} \hat{a} + \frac{\hbar\omega_{eg}}{2}\hat{\sigma}_z + \hbar\Omega_{\rm R}(\hat{a} + \hat{a}^{\dag})\hat{\sigma}_x.
\end{equation}
Since $\hat{\sigma}_x = \hat{\sigma}_- + \hat{\sigma}_+$, the interaction of the Rabi Hamiltonian contains terms proportional to $\hat{a}^{\dag} \hat{\sigma}_+$ and $\hat{a} \hat{\sigma}_-$. These terms are not of the flip-flop type, as they involve the simultaneous creation (destruction) of a photon and an atomic excitation, breaking the intuition based on the absorption/emission paradigm. 
They are called \emph{counter-rotating} because switching to the interaction picture Hamiltonian they evolve as $\hat{a}\hat{\sigma}_-e^{-i(\omega_c +\omega_{eg})t}$, contrary to the flip-flop terms evolving as $\hat{a}\hat{\sigma}_+ e^{-i(\omega_c - \omega_{eg})t}$ \cite{cohen-tannoudji_lagrangian_1997}.
At resonance $\omega_c = \omega_{eg}$, the flip-flop terms become time-independent, while the counter rotating terms keep oscillating with frequency $2\omega_c$~\cite{cohen-tannoudji_lagrangian_1997}.
These terms thus couple states with different energies, and their impact scales in perturbation theory with powers of the coupling $\Omega_{\mathrm{R}}$ divided the bare frequencies $\omega_c$ and $\omega_{eg}$, becoming non-negligible only in the USC regime.

\subsection{The validity of the RWA and the boundary of the USC regime}

In Sec. \ref{sec: Derivation of the JCM} we have seen that the JCM has an internal $U(1)$ symmetry that reduces its Hamiltonian in a block diagonal form composed by a scalar and infinite two-by-two blocks, allowing for a simple analytical solution.
On the contrary the Rabi model does not have this symmetry: the counter-rotating terms do not commute with the excitation number operator $[\hat N, \hat{a}^{\dag}\hat{\sigma}_+]\neq 0$, $[\hat N, \hat{a} \hat{\sigma}_-]\neq 0$. 
Counter-rotating terms, adding or subtracting pairs of excitations, couple only states with even or odd excitation numbers.
The system thus still possesses a $\mathbb{Z}_2$ symmetry, given by the invariance under parity transformation $\hat{a}\mapsto-\hat{a}$, $\hat{\sigma}_-\mapsto -\hat{\sigma}_-$.  
The simple analytical solution of the JCM is not available anymore, even though an exact analytical solution can still be obtained by exploiting the remaining symmetry \cite{braak_integrability_2011}. The structure of the solution is however more intricated and the intuition developed in the JCM is lost \cite{casanova_deep_2010, wolf_dynamical_2013, forn-diaz_broken_2016, stefano_feynman-diagrams_2017}.
In any case, the Rabi model can be easily diagonalized numerically, for instance using the QuTip python library \cite{johansson_qutip_2013}.

\begin{figure}
    \centering
    \includegraphics{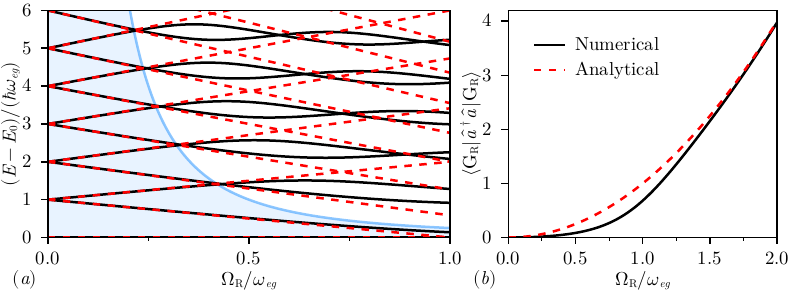}
    \caption{(a) Energy spectrum of the resonant Rabi model (black solid lines) and JCM (red dashed lines) as a function of the coupling strength. The region of validity of the RWA is highlighted in cyan, whose boundary is given by the cyan solid line. (b) The plot of the number of photons in the ground state of the quantum Rabi model by numerical diagonalization (black solid line) and using the first order term from \cref{eq:photon_in_GR} (red dashed line).}
    \label{fig:rwa_break_vacuum}
\end{figure}

The transition from JCM to Rabi model is not sharp, but rather a crossover, where the spectrum of the Rabi model is indistinguishable from the JCM spectrum when the light-matter coupling is sufficiently small $\Omega_{\rm R} \ll \omega_c,\omega_{eg}$, as visible in the left-side of Fig. \ref{fig:rwa_break_vacuum}(a). 
By further increasing the light-matter coupling the difference between the Rabi model and the JCM becomes more prominent due to the increasing importance of the counter-rotating terms.
As pointed out in Ref. \cite{rossatto_spectral_2017}, their relevance can be evaluated by means of perturbation theory by computing the matrix element of the operators $\Omega_{\rm R} \hat{a}\hat{\sigma}_-$ between different $n, n'$ excitation blocks. 
For instance, taking the simplest case of $n, n'=n + 1$, the RWA can be applied only if $\Omega_{\rm R}^2|\langle{n, g|\hat{a}\hat{\sigma}_-| n + 1, e  \rangle}|^2 / (\omega_c+\omega_{eg})^2 \ll 1$.
% \begin{equation}
%     \frac{\Omega_{\rm R}^2}{ 4\omega_c^2 }|\langle{n, g|\hat{a}\hat{\sigma}_-| n + 1, e  \rangle}|^2 < 1,
% \end{equation}
%assuming $\omega_{eg} = \omega_c$.
The RWA regime of validity, delimited by the upper bound of this equation, namely
\begin{equation}
    n_{\mathrm{max}} = \frac{(\omega_c+\omega_{eg})^2}{\Omega_{\rm R}^2},
\end{equation}
is shaded in blue in Fig. \ref{fig:rwa_break_vacuum}(a).
From here is clear that the JCM is a \emph{low-coupling} and \emph{low-energy} effective model of the Rabi one.
Reaching the boundary of validity of the RWA has also well visible observable physical consequences. The most striking one is probably the Bloch-Siegert shift, which was first measured in superconducting circuits \cite{forn-diaz_observation_2010} from the transmission spectroscopy, and successively confirmed also in other different CQED platforms such as Landau polariton setups \cite{li_vacuum_2018}.
Let's note that the breaking of the RWA is the original definition of the USC \cite{ciuti_quantum_2005}, and it was used for its first experimental validation \cite{anappara_signatures_2009}.

\subsection{The non-empty vacuum beyond RWA}

One of the most striking and investigated consequences of the presence of counter-rotating terms is that the empty cavity vacuum (i.e., the JCM ground state $|{\rm G}_{\rm JC}\rangle=|0,g\rangle$) is not anymore an eigenstate of the system.
In fact, in the Rabi model, the ground state coincides with the empty cavity vacuum only for vanishing light-matter coupling and is then progressively filled by photons as the light-matter coupling increases.  
Using the same perturbative approach as in the previous section, one can compute the ground state virtual photon population, obtaining
\begin{align}
\label{eq:photon_in_GR}
 \langle{{\rm G}_{\rm R}| \hat{a}^{\dag} \hat{a} | {\rm G}_{\rm R} \rangle} \propto \left(\frac{\Omega_{\rm R}}{\omega_c} \right)^2 +O\left[\left(\frac{\Omega_{\rm R}}{\omega_c} \right)^4\right].
\end{align}
This result is not specific to the Rabi model only, but rather a generic feature of USC systems. For instance in \figpanel{fig:rwa_break_vacuum}{b} we see the vacuum photon expectation number computed in the quantum Rabi model. The JCM would instead result in a photon population strictly vanishing in the ground state.

The presence of these photons in the coupled ground state (often referred to as \emph{virtual photons}) is a typical non-perturbative phenomenon reminiscent, {\it mutatis mutandis}, of the quark-gluon gas populating the non-empty vacuum of quantum chromodynamics (QCD) \cite{shuryak_qcd_1988}. The possibility of detecting these virtual photons was a central interest in the initial development of the theory of the USC regime \cite{ciuti_quantum_2005}, where CQED systems were identified as ideal playgrounds to study the fascinating and still mysterious physics of vacuum phenomena in QED \cite{milonni_quantum_1994}.
%Differently from free space QED systems (or QCD systems) here is possible to have full control of all the degrees of freedom, designing each setups in such a way to highlight and isolate all quantum phenomena. 

As it will be explained in more detail in the next Sec. \ref{SecOpen} observing quantum vacuum physics is not an easy task, because virtual photons (and virtual particles in general) are normally directly unobservable in experiments and the only proposal in this direction has been to measure the static charge displacement caused by virtual electronic excitations in an asymmetric system \cite{wang_theoretical_2021}. 
Otherwise, the observation of virtual photons is usually studied under a time-dependent perturbation \cite{de_liberato_extracavity_2009} or electrical current \cite{cirio_ground_2016} providing the required energy to convert virtual photons to real ones, or weakly coupling the system to another external probe system and observing its modified emission properties \cite{lolli_ancillary_2015,de_bernardis_light-matter_2023, minganti_phonon_2023}.
%This approach is reminiscent of the \emph{dynamical Casimir} emission and it is another central topic in the development of the USC regime \cite{wilson_observation_2011, hoeb_amplification_2017, dodonov_antidynamical_2017}.

\subsection{Excited states beyond RWA: multi-photon non-linear processes}

The action of the counter-rotating terms in the USC regime impacts also the qualitative nature of the excitation spectrum.
In particular, breaking the conservation of the excitation number allows non-linear processes with the absorption-emission of multiple photons at the same time \cite{ma_three-photon_2015}. This phenomenology is even richer when we explicitly break also the remaining $\mathbb{Z}_2$ symmetry, for instance considering the co-called \emph{asymmetric Rabi model}
\begin{equation}
    \hat{H}_{\rm aR} = \omega_c \hat{a}^{\dag} \hat{a} + \frac{\omega_{eg}}{2}\hat{\sigma}_z + \Omega_{\rm R}(\hat{a} + \hat{a}^{\dag})\hat{\sigma}_x + \frac{\epsilon}{2}\hat{\sigma}_x.
\end{equation}
Here $\epsilon$ quantifies the explicit symmetry breaking and can be interpreted as an external static electric or magnetic field, as commonly employed in circuit QED \cite{yoshihara_superconducting_2016,blais_circuit_2021}.

Increasing the value of $\epsilon$ opens some non-trivial avoided crossings in the spectrum, allowing the system to undergo multi-photon Rabi oscillations that might be used for the generation of multi-photons Fock states \cite{garziano_multiphoton_2015,ma_three-photon_2015,kockum_deterministic_2017}. Here a single atom can emit simultaneously multiple photons with a single transition due to the USC.
Differently from multi-photon processes arising in devices with non-linear quadrupolar light-matter coupling \cite{felicetti_two-photon_2018, felicetti_ultrastrong-coupling_2018}, this dynamics is a USC consequence of the interplay between the linear dipole coupling and the non-linearity (or saturability) of the TLS.
These effects are typically called \emph{tunneling resonances} \cite{de_bernardis_relaxation_2023} since their mathematical description is identical to tunneling resonances appearing in the physics of electronic transport assisted by phonons through molecular or nanostructure quantum dots \cite{koch_theory_2006, leturcq_franckcondon_2009, cui_phonon-mediated_2015, vdovin_phonon-assisted_2016}.
They are currently a major reason of interest in the development of these platforms projecting also new technological perspectives on the USC regime, {\color{black} in terms of new devices for parametric up/down conversion and multi-photon Fock state preparation} \cite{sanchez-burillo_single_2019,macri_spontaneous_2022,koshino_deterministic_2022,mehta_theory_2022, mehta_down-conversion_2023,wang_strong_2024}.

\section{Gauging effective models}
\label{sec:gauging}

When building a theoretical model of a quantum system, it is often useful to consider only a limited subspace of its full Hilbert space (generally the lowest-energy states). In the case of light-matter coupled systems, this projection could introduce the risk of compromising the gauge invariance~\cite{lamb_fine_1952, bassani_choice_1977, de_bernardis_breakdown_2018, stokes_gauge_2019, di_stefano_resolution_2019, taylor_resolution_2020, stokes_implications_2022, arwas_metrics_2023}. Gauge invariance is a fundamental property of QED, ensuring that the dynamics remain unaltered upon gauge transformations. Truncating the Hilbert space can nevertheless break this fundamental symmetry, leading to physical results that depend on the (unphysical) choice of the gauge used to describe the electromagnetic field. 

To better understand the problem, let's consider the two most used gauges to study non-relativistic light-matter interactions: the Coulomb and the Power-Zienau-Woolley gauges~\cite{babiker_derivation_1983,cohen-tannoudji_lagrangian_1997}, where the latter is also known as multipolar or dipole gauge. 
Both gauges are also dubbed in the literature as the $\hat{\mathbf{p}} \cdot \hat{\mathbf{A}}$ or velocity gauge and the $\hat{\mathbf{d}} \cdot \hat{\mathbf{E}}$ or length gauge. Here $\hat{\mathbf{p}}$ and $\hat{\mathbf{d} }= q\hat{\mathbf{x}}$ are the particle momentum and electric dipole moment with charge $q$ and displacement $\hat{\mathbf{x}}$. $\hat{\mathbf{A}}$ and $\hat{\mathbf{E}}$ are the vector potential and the electric field. 
{\color{black} Notice that in many text-books is conventionally adopted the dipole gauge in the derivation of the JCM, motivating our choice in Eq.~\eqref{eq: general form of d * E interaction} instead of reporting the dipole coupling in terms of Coulomb gauge coupling.}
%This arises from the specific form of the interaction term in the Hamiltonian, which will be shown later. 
% It has been already known that these two gauges can lead to different results when applied on subspaces of the full Hilbert space \cite{lamb_fine_1952,bassani_choice_1977, de_bernardis_breakdown_2018}, with the dipole gauge much less affected by the truncation than the Coulomb gauge.
If we consider the transition matrix elements between two states $|j\rangle$ and $|k\rangle$ of a particle with mass $m$, position $\hat{\mathbf{x}}$, and momentum $\hat{\mathbf{p}}$ we can easily derive the equation~\cite{sakurai_modern_2021} 
\begin{align}
\langle j \vert \hat{\mathbf{p}} \vert k \rangle = i m \omega_{jk} \langle j \vert \hat{\mathbf{x}} \vert k \rangle.  
\end{align}
This relation shows that the matrix elements of the momentum are proportional to those of the position, with a proportionality factor linear in the energy difference between the two states $\omega_{jk}$. In the Coulomb gauge, where the interaction Hamiltonian is of the form $\hat{\mathbf{p}} \cdot \hat{\mathbf{A}}$, the matrix element thus vanishes more slowly
with the detuning than in the $\hat{\mathbf{d}} \cdot \hat{\mathbf{E}}$ case, making the approximation of modeling the matter system as a TLS, ignoring out-of-resonance states, more fragile. 

To obtain a quantitative understanding of how serious this problem of gauge non-invariance can be, we can start from 
the Hamiltonian describing a one-dimensional particle interacting with a single photonic mode in the Coulomb gauge $\nabla \cdot \hat{\mathbf{A}} = 0$ (for major details on its derivation see, e.g., Refs. \cite{babiker_derivation_1983,cohen-tannoudji_lagrangian_1997,de_bernardis_cavity_2018})
\begin{equation}
    \label{eq: full Hamiltonian in the Coulomb gauge}
    \hat{\mathcal{H}}_\mathrm{C} = \frac{1}{2 m} \left( \hat{{p}} - q \hat{{A}} \right)^2 + V(\hat{{x}}) + \hbar \omega_c \hat{a}^\dagger \hat{a} \, .
\end{equation}

\begin{figure}[t]
    \centering
    \includegraphics{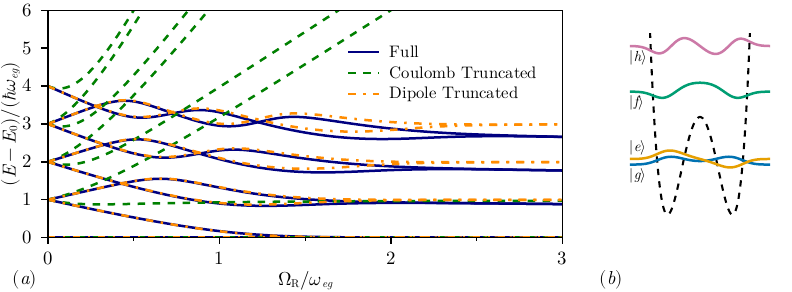}
    \caption{(a) Comparison of the lowest eigenvalues with respect to the ground state energy $E_0$ as a function of the coupling strength $\Omega_\mathrm{R}$, for the two truncated Hamiltonians in the Coulomb ($\hat{H}_C$, green dashed) and dipole ($\hat{H}_D$, orange dash-dotted) gauges. They are compared to the full Hamiltonian $\hat{\mathcal{H}}_\mathrm{C}$ in \cref{eq: full Hamiltonian in the Coulomb gauge}. The Coulomb gauge already deviates at very small values of $\Omega_\mathrm{R}$. In contrast, the dipole gauge breaks down at much higher values. (b) First four eigenstates of the atom Hamiltonian, with the same parameters used in panel (a). The double-well potential is modeled as $V(x) = A x^4 - B x^2$, and the used parameters are $m=1$, $A=50$, and the anharmonicity $ m B^3 / (\hbar^2 A^2) = 45$.}
    \label{fig: gauge_comparison}
\end{figure}

The Hamiltonian in the dipole gauge can be obtained by performing a gauge transformation \cite{craig_molecular_1998, babiker_derivation_1983, cohen-tannoudji_lagrangian_1997}, which is implemented in quantum mechanics by the unitary transformation: $\hat{\mathcal{H}}_\mathrm{D} = \hat{\mathcal{U}} \hat{\mathcal{H}}_\mathrm{C} \hat{\mathcal{U}}^{\dag}$, with
$\hat{\mathcal{U}} = {\exp} [ \frac{q}{\hbar} {A}_0 \hat{{x}} \left( \hat{a} - \hat{a}^\dagger \right)]$, and $A_0$ the zero point fluctuation of the vector potential. The two Hamiltonians $\hat{\mathcal{H}}_\mathrm{C}$ and $\hat{\mathcal{H}}_\mathrm{D}$ have the same spectrum, being related by a unitary transformation. 
Projecting them onto the two lowest-energy states $\ket{g}$ and $\ket{e}$ of the uncoupled matter system, through the projection operator $\hat{P} = \dyad{g} + \dyad{e}$, we obtain the Hamiltonians describing the interaction of a single-mode electromagnetic field with a two-level system: $\hat{H}_\mathrm{C} = \hat{P} \hat{\mathcal{H}}_\mathrm{C} \hat{P}$ in the Coulomb gauge, and $\hat{H}_\mathrm{D} = \hat{P} \hat{\mathcal{H}}_\mathrm{D} \hat{P}$ in the dipole gauge.

In \figpanel{fig: gauge_comparison}{a} we compare the eigenvalues of the full Hamiltonian in \cref{eq: full Hamiltonian in the Coulomb gauge} with those of the of two projected Hamiltonians $\hat{H}_\mathrm{C}$ and $\hat{H}_\mathrm{D}$ as a function of the vacuum Rabi frequency $\Omega_\mathrm{R} \equiv q \omega_c A_0 \inlinemel{e}{\hat{x}}{g} / \hbar$. We consider a quartic potential in which the energy separation between the first two modes is 12 times smaller than the gap between the second and the third.
 As the coupling increases, deviations between the three models occur, already at the onset of the USC regime in the Coulomb gauge and for much larger values of the coupling in the dipole gauge. However, increasing the atom's anharmonicity extends the range of agreement of the dipole gauge with the untruncated case.
 It is worth noticing that the gauge choice that reduces the error on the truncated Hilbert space is system dependent~\cite{stokes_gauge_2019, ashida_cavity_2021,ashida_nonperturbative_2022, arwas_metrics_2023}.

Gauge invariance can be also implemented already  in the two-level subspace by performing the minimal coupling replacement directly in the truncated subspace, that is applying the following unitary transformation to the free photon Hamiltonian only ~\cite{di_stefano_resolution_2019, taylor_resolution_2020}
\begin{align}\label{eq:U_polaron_Coulomb}
\hat{U} = {\exp} \left[ \frac{q}{\hbar} A_0 \hat{P} \hat{x} \hat{P} \left( \hat{a} - \hat{a}^\dagger \right)\right] = {\exp} \left[\frac{\Omega_\mathrm{R}}{\omega_c} \hat{\sigma}_x \left( \hat{a} - \hat{a}^\dagger \right)\right].    
\end{align}
{\color{black}While the results are completely equivalent to what was described above, this approach represents the basis of lattice gauge theories~\cite{savasta_gauge_2021}, becoming another interesting example of how cavity QED can be a useful playground to experiment with the most complex concepts of modern physics.
It is worth mentioning that, as much as in other branches of physics \cite{fukuda_gauge_1976, nielsen_gauge_1975}, the discussion about gauge invariance has risen an intense debate in the community, which is still on going~\cite{stefano_reply_2024,stokes_gauge_2024}.
Moreover, while here we focused mainly on Coulomb and dipole gauge, other gauge choices may be also convenient, dependently from the specific problem the one needs to discuss \cite{ashida_cavity_2021}.
}

{\color{black} 
We conclude this section by noticing that both dipole and Coulomb gauge returns a Hamiltonian that contains the counter-rotating terms in the light-matter coupling. However it was shown in Refs. \cite{drummond_unifying_1987,stokes_gauge_2019} that is possible to implement an intermediate gauge transformation which exactly cancels the counter-rotating terms. In this representation the RWA is no longer an approximation and the vacuum is always empty.
While we will come back on the relative meaning of the quantum vacuum in the last section, it is important to understand that, under the standard assumptions given above and described in Ref. \cite{de_bernardis_breakdown_2018}, this representation pays the price of being not fully compatible with the TLA as much as the Coulomb gauge \cite{arwas_metrics_2023}.
}

\section{Single particle vs collective coupling}
%The JCM as well as the Rabi model are characterised by the extreme nonlinearity of their TLS: the system can contain at most one single matter excitation. 
Aside from the modification of the photonic vacuum, the USC is predicted to modify also the state and properties of the matter counterpart, represented in the JCM or Rabi model by a TLS \cite{zueco_qubit-oscillator_2009,casanova_deep_2010,manucharyan_resilience_2017, de_bernardis_relaxation_2023}.
From this observation, a strong interest has arisen to modify and control the properties of electrons, molecules, or devices exploiting the quantum fluctuations of the USC vacuum \cite{garcia-vidal_manipulating_2021,bloch_strongly_2022}.
For instance, the USC between a single electron and the resonator has been shown to have a strong impact on electron transport
\cite{valmorra_vacuum-field-induced_2021}, thus becoming very interesting for more involved device operations with application purposes \cite{lagree_effective_2023,pisani_electronic_2023}.
Generalizing this concept to any light-matter systems is certainly appealing as a powerful technological framework, but also as a new way to explore the fundamental science behind the quantum vacuum \cite{milonni_quantum_1994}.

\subsection{Bosonising the light-matter interactions}
Solid-state CQED setups, especially those in which USC has been achieved, are usually not well described by the JCM nor by the Rabi model because of the presence of multiple dipoles participating in the light-matter dynamics.
A minimal description is provided by generalizing the JCM or the Rabi model where multiple TLSs are identically coupled to the same photonic mode (which means their separation is much smaller than the wavelength), giving the so-called Dicke model \cite{kirton_introduction_2019}
\begin{equation}
    \hat{H}_I \propto \Omega_{\rm R}(\hat{a}+\hat{a}^{\dag})\sum_{i=1}^N\hat{\sigma}_x^i \approx \sqrt{N}\Omega_{\rm R} (\hat{a}+\hat{a}^{\dag})(\hat{b}+\hat{b}^{\dag}).
\end{equation}
Here the index $i$ on the Pauli matrices addresses the different TLSs. 
We introduced the collective annihilation operator $\hat{b} = \sum_{i=1}^N\hat{\sigma}_-^i / \sqrt{N}$, describing a collective excitation of all the TLSs. This operator satisfies $[\hat{b},\hat{b}^{\dag}]\approx 1+O(\frac{n_x}{N})$ in the dilute regime, that is when the number of excitations $n_x$ is much smaller than the total number $N$ of TLSs \cite{holstein_field_1940, klein_boson_1991}. {\color{black} It is worth noticing that in this same limit, the equilibrium Dicke model is exactly solvable~\cite{hepp_superradiant_1973,wang_phase_1973,roman-roche_effective_2022}, making it a paradigmatic example to discuss basics many-body effects in light-matter interactions. }

This transformation also leads to a vacuum Rabi frequency enhanced by a factor $\sqrt{N}$, often referred as \emph{collective enhancement}, which makes it much simpler to reach extreme values of the coupling \cite{ciuti_quantum_2005, anappara_signatures_2009, todorov_ultrastrong_2010,scalari_ultrastrong_2012}. 
Increasing the number of dipoles the coupling increases but the optical nonlinearity decreases. The saturation of the TLS washes out, as the system is able to absorb multiple photons. 
The system becomes then well described by a linear optical approach, where both the cavity and the material are described by harmonic oscillators. 

\begin{figure}
    \centering
    \includegraphics{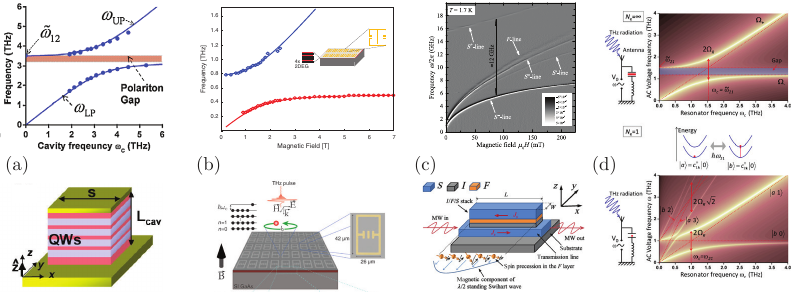}
    \caption{Solid-state CQED setups and their spectral features changing as they approach the single-dipole nonlinear regime.
    (a) ISB polariton experimental setup and transmission spectrum, taken from Ref. \cite{todorov_ultrastrong_2010}. (b) Landau polariton setup and transmission spectrum, taken from Ref. \cite{scalari_ultrastrong_2012}. (c) Magnonic setup and transmission spectrum, taken from Ref. \cite{golovchanskiy_approaching_2021}. (d) Comparison between collective and single particle transmission spectrum, taken from Ref. \cite{todorov_few-electron_2014}. }
    \label{fig:exp_coll_coupling}
\end{figure}

In Fig. \ref{fig:exp_coll_coupling}(a)-(c) we show three paradigmatic examples of CQED setups well described by the bosonic approximation: intersubband polaritons \cite{todorov_ultrastrong_2010}, Landau polaritons \cite{scalari_ultrastrong_2012} and magnonic polaritons \cite{golovchanskiy_approaching_2021}. 
In order to recover the technologically relevant nonlinear regime \cite{cominotti_theory_2023}, there has been a constant effort in those systems to reduce the number of dipoles while keeping the system in the USC regime \cite{todorov_few-electron_2014, keller_few-electron_2017, ballarini_polaritonic_2020,rajabali_ultrastrongly_2022}. Spectral differences at the transition between collective and single particle physics are shown in Fig. \ref{fig:exp_coll_coupling}(d). 
{\color{black} It is important to point out that Bosonized excitations with non-negligible non-linearities are also found in phonon-polaritons THz materials, which is another great area of interest regarding the USC impact on matter~\cite{mazza_superradiant_2019,ashida_quantum_2020, ashida_cavity_2021, ashida_nonperturbative_2022,li_electromagnetic_2020, mazza_collective_2024}.  }

\subsection{Large $N$ problem}

What impact the linearity of the Dicke model has on the possibility of modifying ground-state properties is not immediately clear, given the multitude of possible interactions between the dipolar degree of freedom coupled to the photonic field and all the other internal degrees of freedom of the quantum emitter.
Analyzing different simplified models, it has been shown that many single-particle effects are not enhanced by the collective coupling and scale with zero or negative powers of $N$ \cite{cwik_excitonic_2016,cortese_collective_2017, pilar_thermodynamics_2020}, with adverse consequences on the possibility to modify the equilibrium properties of the matter involved \cite{galego_cavity_2019,martinez-martinez_can_2018, pilar_thermodynamics_2020, kansanen_theory_2023}.

An intuitive explanation behind the lack of an impact of collective USC on the state of a single TLS \cite{cortese_collective_2017} can be grasped by noticing that, to the lowest order, the energy shift of a coupled collective eigenmode is of order $\Delta E\propto \sqrt{N}\Omega_{\rm R}$, and the maximal change in such an energy when the internal state of one molecule changes is of order $\frac{\partial \Delta E}{\partial N}\propto \frac{\Omega_{\rm R}}{\sqrt{N}}$. The force acting on the single dipole is thus vanishing in the thermodynamics limit $N\rightarrow\infty$.

However, some experimental works showed a change in chemical reactions \cite{garcia-vidal_manipulating_2021, nagarajan_chemistry_2021}, a shift in the critical temperature of specific material properties \cite{jarc_cavity-mediated_2023} or a modification of the macroscopic quantum Hall transport properties \cite{appugliese_breakdown_2022} when the system is embedded in a resonant cavity.
These systems have a significant coupling strength only considering their collective coupling, thus contradicting the theoretical predictions cited above.
Other theoretical works have shown that considering more sophisticated and complex models the collective coupling might have a macroscopic effect on the total reaction process or the material properties \cite{schafer_shining_2022, lenk_collective_2022, fadler_engineering_2023}. 
The clash between intuitive results, experimental facts, and complex ab-initio calculations has opened a debate that is still unsolved.

\section{Open quantum systems: can we measure the non-empty vacuum?}
\label{SecOpen}

The photon flux leaking out of a resonator can be usually approximated as the number of photons in the cavity times their escape rate.
The presence of photons in the ground state predicted by the USC regime (see Eq. \ref{eq:photon_in_GR}) immediately shows how such an intuitive picture fails in this non-perturbative regime, which would otherwise predict photonic emission from the ground state, breaking energy conservation.

Multiple approaches have been developed to correctly deal with such an issue ~\cite{ciuti_input-output_2006, de_liberato_extracavity_2009,bamba_system-environment_2013,bamba_recipe_2014,settineri_dissipation_2018,di_stefano_photodetection_2018}. Without getting into the technical details required to understand the subtle differences between these various approaches, we will try here to build an intuition of the problem
with the standard approaches to open quantum systems.
To this aim we will consider the standard Lindblad master equation describing the evolution of the density matrix of a system coupled to a zero temperature reservoir, leading to a loss rate $\gamma$
\begin{equation}
    \label{eq: standard Lindblad superoperator}
    \mathcal{L}_{\mathrm{std}} \hat{\rho} =  \gamma  \mathcal{D} \left[ \hat{S} \right] \hat{\rho},
\end{equation}
where $\hat{S}$ is the operator describing the loss of a bare excitation in the system, the Lindblad dissipator is defined as
\begin{equation}
    \label{eq: Lindblad dissipator}
    \mathcal{D} \left[ \hat{S} \right] \hat{\rho} = \frac{1}{2} \left[ 2 \hat{S} \hat{\rho} \hat{S}^\dagger - \hat{S}^\dagger \hat{S} \hat{\rho} - \hat{\rho} \hat{S}^\dagger \hat{S}  \right],
\end{equation}
and we neglected the part describing unitary evolution.
It is easy to verify that if the ground state is the vacuum for the bare excitations, $\hat{S}\ket{G}=0$, then $\mathcal{L}_{\mathrm{std}}\dyad{G}{G}=0$ and the ground state is stable against losses.
This is the case for example for the JCM ground state when the operator $\hat{S}$ describes a photonic ($\hat{a}$) or matter ($\hat{\sigma}_-$ ) loss
\begin{equation}
\hat{a}\ket{G_{\text{JC}}}=\hat{\sigma}_-\ket{G_{\text{JC}}}=0.    
\end{equation}
This is not the case anymore when the ground state is not the vacuum for the bare excitations $\hat{S}\ket{G}\neq 0$, 
as clearly shown in Eq. \ref{eq:photon_in_GR} for the Rabi model. This leads to $\mathcal{L}_{\mathrm{std}}\dyad{G_{\text{R}}}{G_{\text{R}}}\neq0$ and the ground state is unstable against losses.

The issue lies in the details of the master equation derivation, which is derived using the bare basis ($ \{g, e \} \otimes \{n \in \mathbb{N} \} $) rather than in the energy eigenbasis, and assumes a white reservoir whose density of states can be considered constant in the spectral interval of reference. While these are usually safe approximations, they catastrophically fail in the USC regime, where the energy shifts are of the same order as the bare frequencies. In these conditions, a fully white reservoir implicitly assumes to have a non-vanishing density of states also at negative energies, which explains the instability of the ground state as the emission of unphysical negative-energy excitations in the reservoir.

\begin{figure}[t]
    \centering
    \includegraphics{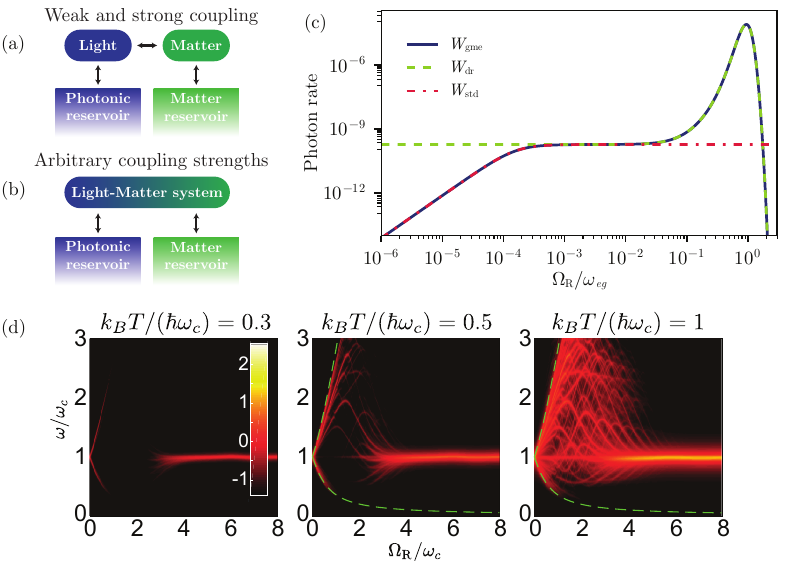}
    \caption{(a) Schematic description of an open hybrid system whose subparts are in weak interaction. (b) The same as before, but this time the two subsystems are ultrastrongly coupled, and the model of the interaction with the environment must take into account their hybridization. (c) Comparison of three different methods to describe open quantum systems, showing the photon emission rate at the steady-state as a function of the vacuum Rabi frequency $\Omega_\mathrm{R}$. The generalized master equation is the only one that works across all values of the coupling. (d) The logscale intensity of the spectrally resolved black-body radiation emitted from a cavity QED system as a function of the light-matter coupling. While for intermediate couplings the two polaritonic branches are well visible, deep in the USC regime the spectrum collapses to the uncoupled cavity emission spectrum. Figure taken from Ref. \cite{pilar_thermodynamics_2020}. }
    \label{fig: master equation}
\end{figure}

This problem can be solved by not performing the white-reservoir approximation and deriving the Liouvillian in the eigenbasis of the light-matter Hamiltonian~\cite{ciuti_input-output_2006,beaudoin_dissipation_2011}. The resulting \emph{dressed}
master equation takes the form
\begin{equation}
    \begin{split}
        \label{eq: dressed Lindblad superoperator}
        \mathcal{L}' \hat{\rho} =&  \sum_{j,k} \gamma(\omega_{jk})   \mathcal{D} \left[ \dyad{k}{j} \right] \hat{\rho},
    \end{split}
\end{equation}
where the dissipator is expressed in terms of jump operators between the $k$-th and the $j$-th eigenstates of the coupled light-matter Hamiltonian, $\dyad{k}{j}$, with the associated Bohr frequency $\omega_{jk}=\omega_j - \omega_k$.
{\color{black}Consistently with the secular approximation (or just RWA) needed to obtain the master equation in a Linblad form, one needs to discard all the transitions with negative frequencies that simultaneously create an excitation in the bath, for which in the above expression we must enforce that $j>k$. 
The problem of energy conservation is now solved, being only an artifact of a wrong derivation.
As a consequence, the open dynamics must be interpreted as transitions between the true eigenstates, where light and matter are entangled together and it is not possible to simply distinguish them separately, as schematized in Fig. \ref{fig: master equation}(a-b)
The above derivation is well justified at weak system-bath coupling and it can be equivalently rephrased as a vanishing density of states of the bath at negative frequencies leading to
$\gamma(\omega<0)=0$. }
%However this view could be still problematic since it does not clearly account the need of a small system-bath coupling.

{\color{black} It is worth noticing that the above approach relies on having a different bath for each transition \cite{beaudoin_dissipation_2011}. This remains true even if the all the system's transitions decay in the same bath, provided that their coupling to the bath is smaller than their relative distance in frequencies. In the weak coupling regime of cavity QED for instance this condition is removed and} a problem with this formulation can nevertheless emerge due to the degenerate nature of the spectrum in such a regime. This in turn led to the derivation of \emph{generalized} master equations ~\cite{settineri_dissipation_2018} explicitly taking these degeneracies into account and leading to correct results for all values of the light-matter coupling. {\color{black}Similar issues are partially discussed also in Refs. \cite{zueco_ultrastrongly_2019,lednev_lindblad_2024, de_la_pradilla_taming_2024} and, more completely in a very extensive way, in Ref. \cite{cattaneo_local_2019}.
}

Although the dressed and the generalized master equations at zero temperature lead the system to the ground state, the detection of each quantity can still lead to major mistakes due to the possibly wrong representation of the observables.
As a concrete example, we show how the photodetection has to be revised in the USC. 
Indeed, we already mentioned that $\langle \mathrm{G}_\mathrm{R} \vert \hat{a}^\dagger \hat{a} \vert \mathrm{G}_\mathrm{R} \rangle \neq 0$. This suggests that $\hat{a}^\dagger \hat{a}$ can no longer be interpreted as the output photon rate, otherwise, we would have photon emission even in the ground state. Following the standard theory of photodetection~\cite{glauber_quantum_1963, cohentannoudji_atomphoton_1998, walls_quantum_2008}, the output photon rate for a light-matter system in the generic state $|\Psi \rangle$ is proportional to
\begin{equation}
\label{eq:photon-rate}
    W \propto \mel{\Psi}{\hat{E}^- \hat{E}^+}{\Psi} \, ,
\end{equation}
where $\hat{E}^{+}$ ($\hat{E}^- = (\hat{E}^{+})^\dagger$) is the positive (negative) frequency part of the electric field operator. In case of weak coupling we have $\hat{E}^+ \propto \hat{a}$ and we obtain the usual formula. In the USC regime, however, we have a hybridization of the frequencies and the positive frequency operator becomes
\begin{equation}
    \hat{E}^+ = \sum_{k>j} \mel{j}{\hat{E}}{k} \dyad{j}{k} \, ,
\end{equation}
which gives $\langle \mathrm{G}_\mathrm{R} \vert \hat{E}^- \hat{E}^+ \vert \mathrm{G}_\mathrm{G} \rangle = 0$. Moreover, the form of the electric field operator becomes gauge dependent~\cite{pilar_thermodynamics_2020,settineri_gauge_2021,mercurio_regimes_2022}, but leaving the photon rate in \cref{eq:photon-rate} gauge-invariant. {\color{black} Similar analysis was carried out on pure dephasing processes, where the system experienced a loss of coherence due to stochastic fluctuations of the bare energies of the systems~\cite{beaudoin_dissipation_2011,mercurio_pure_2023}. 
}

Aside from the correct formalism to adopt, these issues make us reflect on the meaning of photons in the presence of matter, a point that was raised already a while ago \cite{lamb_anti-photon_1995}.
Virtual photons arising from the USC to matter cannot be simply interpreted as quanta of the transverse electric field oscillations, as commonly done in the free space case. Their physical meaning depends on the gauge that we adopt to describe the interaction with matter and it must be handled with care \cite{de_bernardis_cavity_2018,stokes_gauge_2019, rouse_theory_2023, stokes_gauge-relativity_2023}.

In order to provide a concrete example of the formalism introduced here, in Fig. \ref{fig: master equation}(c) we show results for the rate of emitted photons out of a Rabi model where both the cavity and the two-level atom are in interaction with their respective environment. 
The cavity reservoir is at zero temperature ($T_c = 0$), while that of the atom has a relatively low temperature $T_a = 0.05 \hbar \omega_{eg} / k_\mathrm{B}$. The atom environment is the only source of energy (thermal pumping), and thus the emitted photons are only the result of interaction with the atom.
The results are shown for the standard (std), dressed (dr), and generalized (gme) master equations, clearly showing their respective regions of applicability.
The well-known Purcell effect~\cite{purcell_resonance_1946} is clearly visible in the weak coupling region, where the photon emission rate increases with $\Omega_\mathrm{R}$. 
The signature of the USC can be related to the increase in the photon emission of several orders of magnitude, while the sudden decrease is related to the decoupling effect of the deepstrong coupling regime~\cite{de_liberato_light-matter_2014,garcia-ripoll_light-matter_2015,mueller_deep_2020,ashida_cavity_2021, de_bernardis_relaxation_2023}, which will be discussed in details in the next section.
In Fig. \ref{fig: master equation}(d) we show the black-body emitted spectrum derived in Ref. \cite{pilar_thermodynamics_2020} within the dressed framework. At low temperatures, for intermediate couplings, the emitted thermal radiation shows the presence of the two polaritonic branches that collapse in a single line at higher coupling strength. Also here we see another manifestation of the non-perturbative light-matter decoupling effect. At higher temperatures, other lines appear at intermediate couplings, before the light-matter decoupling regime, due to non-linear multi-photon transitions, similar to the single-particle transmission spectra from Ref. \cite{todorov_few-electron_2014} shown in Fig. \ref{fig:exp_coll_coupling}(d).

\section{Recovering the JCM in the deepstrong coupling regime: the light-matter decoupling}
\label{SecDecoupling}

After having explored the different ways in which the JCM can break for sufficiently large values of the light-matter coupling, in this last Section we will close the loop, showing how a JCM can be recovered can be recovered for even stronger interaction strengths.
 
\subsection{Light-matter decoupling}
One surprising phenomenon of non-perturbative CQED is the so-called \emph{light-matter decoupling}: while increasing the coupling strength between light and matter typically makes their dynamics more correlated and entangled, for $\Omega_{\rm R}>\omega_c,\omega_{eg}$
this trend is reversed, and light and matter are rapidly decoupled.
As already anticipated in the previous section this feature is well visible in Fig. \ref{fig: master equation} where the photonemission rate drops to zero at large coupling strength.

\begin{figure}
    \centering
    \includegraphics{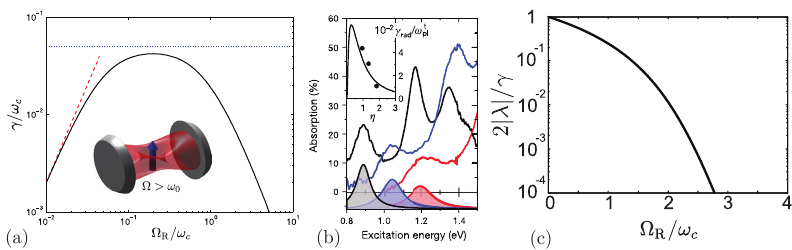}
    \caption{(a) Theoretically predicted electroluminescence emission rate as a function of the coupling strength, taken from Ref. \cite{de_liberato_light-matter_2014}. (b) Experimentally measured absorption spectra of a plasmonic nanoparticle setup in the USC regime and (in the inset) its radiative damping rate as a function of the light-matter coupling strength, taken from Ref. \cite{mueller_deep_2020}. (c) Liouvillian gap of the open-dissipative Rabi model as a function of the coupling strength, taken from \cite{de_bernardis_relaxation_2023}. }
    \label{fig: LM_decoupling}
\end{figure}

The light-matter decoupling was first reported in Ref. \cite{de_liberato_light-matter_2014} in the context of harmonic polariton systems where it was interpreted as a metallization of the optical response of a dielectric for extreme values of the dipolar moment.
The dipoles then become perfect metallic mirrors and expel the electromagnetic field (inset in Fig. \ref{fig: LM_decoupling}(a) ).
This first prediction was indirectly confirmed in Ref. \cite{bayer_terahertz_2017}.
One important prediction of Ref. \cite{de_liberato_light-matter_2014}, shown in Fig. \ref{fig: LM_decoupling}(a)  is that the Purcell effect \cite{purcell_resonance_1946} breaks down in the USC regime, and the spontaneous emission rate changes non-monotonically with the light-matter coupling, an effect experimentally measured in Ref. \cite{mueller_deep_2020} and reported in Fig. \ref{fig: LM_decoupling}(b).

The USC light-matter decoupling was also investigated in the context of superconducting circuit QED, for instance in Ref. \cite{garcia-ripoll_light-matter_2015, jaako_ultrastrong-coupling_2016}, exhibiting a very similar phenomenology.
Another related consequence was illustrated and generalized in Ref. \cite{de_bernardis_relaxation_2023}, where is predicted an exponential slow-down in the thermalization and relaxation dynamics of any USC system. 
Its quantification can be deduced by the Liouvillian gap \cite{minganti_spectral_2018} of the USC master equation described in the previous section (see Fig. \ref{fig: LM_decoupling}(c)).

\subsection{The gRWA and the polaronic JCM}
The origin of the USC light-matter decoupling in single-mode systems can be traced back to the existence of a hidden approximation: the generalized rotating-wave approximation (gRWA).
%While it is clear that the USC phenomenology of the Rabi model appears much more complicated and vast than the simple absorption/emission dynamics described by the JCM, at a closer look 
As it was first reported in Ref. \cite{irish_generalized_2007} and then developed in Ref. \cite{yu_analytical_2012}, it is possible to recover the simple physics of the JCM when the coupling strength places the system deep into the USC regime, where $\Omega_{\rm R}/\omega_{c}\gg 1$.
As a consequence, and apparently contradicting what we reported in the previous Sections, the ground state of the system appears as a trivial empty vacuum.

%This can seem in to contradict with all that we have exposed so far and even with the main motivations moving the initial research in the USC regime but it will be soon clear that there is no contradiction at all.

%This seems in strong contradiction with what we have exposed so far and indeed it requires further elaboration.

%So far we have introduced a picture of the USC regime which could not differ more than the usual JCM picture:
%the spectrum of the light-matter coupled system becomes completely different \cite{rossatto_spectral_2017, frisk_kockum_ultrastrong_2019} and several standard approximations break when we enter in the non-perturbative regime \cite{de_bernardis_cavity_2018};
%even the relation with dissipations and external reservoirs is strongly affected by the USC \cite{beaudoin_dissipation_2011, forn-diaz_ultrastrong_2019}.
%However this strong difference between standard JCM regime and USC regime is only apparent, and, behind the formalism things look much more similar than expected.
%Indeed, it was first realised in Ref. \cite{irish_generalized_2007} and then recaptured in Ref. \cite{yu_analytical_2012}, that also the Rabi model in USC is mostly based on these concepts of absorption and emission introduced in Sec. \ref{sec: Derivation of the JCM}.
%Differently to the JCM, this conceptual simplicity is hidden in the mathematical structure of the Rabi model eigenstates and it can only be pulled out by introducing the generalized rotating-wave approximation (gRWA).

The gRWA is not the standard RWA that leads to the JCM because it first requires transforming the Rabi Hamiltonian in Eq. \eqref{eq:ham_Rabi} in a new basis, often called \emph{polaron frame}~\cite{nazir_ground_2012,bera_generalized_2014}.
The coordinate transformation is implemented by the same unitary transformation in Eq. \eqref{eq:U_polaron_Coulomb} which was used to implement the minimal substitution in the Coulomb gauge Hamiltonian directly in the truncated two-level subspace.
In this context, unrelated to gauge choices, this transformation is known as \emph{polaron transformation}.
The polaron frame Rabi Hamiltonian (which coincides with the Coulomb gauge TLS Hamiltonian presented at the end of Sec. \ref{sec:gauging}) is then given by
\begin{equation}\label{eq:ham_Rabi_polaron}
    \hat{H}_{\rm R}^{\rm pol} = \hat{U}^{\dag}\hat{H}_{\rm R} \hat{U} = \omega_c \hat{a}^{\dag} \hat{a} + \frac{\omega_{eg}}{2}\left[\cos \hat{\theta} \hat{\sigma}_z - \sin \hat{\theta} \hat{\sigma}_y\right],
\end{equation}
where $\hat{\theta}= -i2\Omega_{\rm R}/\omega_c(\hat{a} - \hat{a}^{\dag})$.
At this point, one can Taylor expand the trigonometric operators considering that $\cos \hat \theta = (\hat{U}_D(\Omega_{\rm R}/\omega_c) + \hat{U}_D(\Omega_{\rm R}/\omega_c)^{\dag})/2$ and $\sin \hat \theta = -i(\hat{U}_D(\Omega_{\rm R}/\omega_c) - \hat{U}_D(\Omega_{\rm R}/\omega_c)^{\dag})/2$ \cite{irish_generalized_2007,yu_analytical_2012, de_bernardis_relaxation_2023}. 
These operators are expressed in terms of displacement operators $\hat{U}_D = \exp [ (x^*\hat{a} - x\hat{a}^{\dag}) ]$, allowing for a normal order expansion \cite{cahill_ordered_1969} that permits to easily isolate the positive/negative frequency contributions.
After some tedious but straightforward passages, one can re-express the interaction Hamiltonian in a JCM-like form \cite{irish_generalized_2007, yu_analytical_2012, de_bernardis_relaxation_2023}
\begin{equation}
 \hat{H}_{\rm R}^{\rm pol}\approx\, \omega_{eg}e^{-2\Omega_{\rm R}^2/\omega_c^2}\left[ f(\hat{a}^{\dag}\hat{a}) \hat{a} \hat{\sigma}_+ + f^*(\hat{a}^{\dag}\hat{a})\hat{a}^{\dag} \hat{\sigma}_-\right],   
\end{equation}
where $f$ is a complicated polynomial function \cite{irish_generalized_2007}.
This expression is still mathematically very complicated but has the advantage of clearly showing that also the Rabi model is mostly built over the concept of absorption/emission.
The exponential coefficient in front $e^{-2\Omega_{\rm R}^2/\omega_c^2}$ {\color{black}can be seen as a non-perturbative Lamb-shift \cite{diaz-camacho_dynamical_2016, forn-diaz_ultrastrong_2017}}, emerging also as a feature of the polaron transformation, and explicitly accounts for the light-matter decoupling in the infinite coupling limit $\Omega_{\rm R} \rightarrow \infty$.
It also explains why the gRWA is applicable in such a limit, suppressing the polaron light-matter interaction term which is then treated in perturbation theory, similarly to what is introduced in Sec. \ref{sec: Derivation of the JCM}.

As a specific property, the ground state in the polaron frame is given by $|{\rm G}_{\rm pol}\rangle \approx |0, g\rangle$ being indeed the empty cavity vacuum. 
To obtain the ground state in the standard frame (or Rabi vacuum) we have to transform this state back, obtaining
\begin{equation}
    |{\rm G}_{\rm R}\rangle \approx \hat{U} |0, g\rangle  = \frac{1}{\sqrt{2}}\left( |\alpha, \rightarrow\rangle + |-\alpha, \leftarrow\rangle \right),
\end{equation}
where $|\alpha\rangle$ is a photon's coherent state with amplitude $\alpha = \Omega_{\rm R}/\omega_c$ and $|\leftarrow / \rightarrow\rangle = (|e\rangle \pm |g\rangle )/\sqrt{2}$.
We then notice that all the notions of \emph{vacuum} and \emph{virtual photons} are only relative to the specific considered frame (or Hilbert space basis). Moreover, we see that the Rabi vacuum is not only non-empty but is also highly entangled since for $\Omega_{\rm R} \gg \omega_c$ it is a \emph{cat state} \cite{haroche_nobel_2013}.
In the polaron basis, the ground state instead becomes quite trivial and loses all the entanglement (which is also a relative property).
This observation stimulated the idea that in light-matter problems one can always find a \emph{disentangling} transformation that strongly simplifies the description.
In particular, this was explored in multimodal light-matter systems, where the polaron transformation is generalized to treat the USC regime on the basis of minimizing the entanglement between matter and light. 
In Ref. \cite{diaz-camacho_dynamical_2016,shi_ultrastrong_2018, zueco_ultrastrongly_2019} is shown that a generalized polaron transformation can be used to extract semi-analytical approximated solutions of the multi-mode USC systems. 

{\color{black}
To close this section, we notice that the existence of a disentangling transformation such as the polaron one does not immediately imply the aforementioned light-matter decoupling since that is a property of the full dynamics and not just of the groundstate. However, in the contest, for instance, of the Rabi model, it realizes an optimal basis from which the light-matter decoupling phenomenon emerges quite clearly. In this sense one should interpret the implication in a reversed way: because of the light-matter decoupling a disentanged basis offers a simpler and more natural view of the system.
In this perspective, while light-matter decoupling exists on every basis, its manifestation in formalism can be very different from one representation to another. 
For instance, it was recently shown~\cite{ashida_cavity_2021} that considering a different frame than the polaron one the light-matter decoupling can be seen as the increasing of the electron effective mass $m_\mathrm{eff} = m [1 + 2 (\Omega_\mathrm{R} / \omega_c)^2]$, giving rise to a tight localization of the electron around the potential minima.
}

\section{Conclusion}
\label{sec:conclusion}

In this paper, we tried to provide a non-technical report on the relevance of the JCM in the contemporary research landscape. The vitality of such a topic of investigation can be easily gauged by the fact that essentially all the approximations on which the JCM rests have been tested, stretched, and broken in one way or another. The JCM remains today an important toy model to approach the topic of light-matter coupling at the quantum level. Still, it is best understood as one node of various related models, mapping a much broader section of the parameter space and hosting a corresponding much-richer phenomenology, which we are sure will continue to fascinate researchers for many years to come.

\begin{backmatter}
\bmsection{Funding}
S.D.L. acknowledges funding from the Leverhulme Trust through the grant RPG-2022-037 and the Philip Leverhulme Prize. D.D.B. acknowledges funding from the European Union - NextGeneration EU, "Integrated infrastructure initiative in Photonic and Quantum Sciences" - I-PHOQS [IR0000016, ID D2B8D520, CUP B53C22001750006].

%\bmsection{Acknowledgments}
%The authors wish to thanks Prof. X.

\bmsection{Disclosures}
The authors declare no conflicts of interest.

\bigskip

\bmsection{Data availability} 
No data were generated or analyzed in the presented research.

\bigskip

%\noindent Data availability statements are not required for preprint submissions.

%\bmsection{Supplemental document}
%See Supplement 1 for supporting content. 

\end{backmatter}

%%%%%%%%%%%%%%%%%%%%%%% References %%%%%%%%%%%%%%%%%%%%%%%%%
\bibliography{references_zotero}

\end{document}